\documentclass[oldversion]{aa}

\usepackage{epsfig}
\usepackage{graphics}
\usepackage{float}
\usepackage{amsmath}
\usepackage{multirow}
\usepackage{longtable}
\usepackage{rotate}
\usepackage{array}
\usepackage{subfigure}
\def\kms{\ifmmode{\rm km\,s^{-1}}\else\hbox{$\rm km\,s^{-1}$}\fi}

\setlongtables

\begin{document}

\title{Bayesian model checking: A comparison of tests}

\author{L.B.Lucy}
\offprints{L.B.Lucy}
\institute{Astrophysics Group, Blackett Laboratory, Imperial College 
London, Prince Consort Road, London SW7 2AZ, UK}
\date{Received ; Accepted }

\abstract{Two procedures for checking Bayesian models are compared
using a simple test problem based on the local Hubble expansion. Over four 
orders of magnitude, $p$-values derived
from a global goodness-of-fit criterion 
for posterior probabiity density functions (Lucy 2017) agree closely with 
posterior
predictive $p$-values. The former can therefore serve as an effective proxy
for the difficult-to-calculate posterior predictive $p$-values. 
}

\authorrunning{Lucy}
\titlerunning{Bayesian models}
\maketitle

\section{Introduction}

In a recent paper (Lucy 2017; L17), straightforward procedures are 
proposed for Bayesian model checking. 

Because all Bayesian inferences 
derive from the posterior probability density function $\Lambda(\alpha|D)$,
where $\alpha$ is the parameter vector and $D$ is the data, these 
tests check the global goodness-of-fit (GOF) to $D$ provided by $\Lambda$.
The proposed checks take the form of $p$-values closely analogous
to those familiar in frequentist analyses. The first test derives from the
$\chi^{2}_{B}$ statistic (L17, Appendix C) for uncorrelated 
measurement errors.
The second test derives from the $\psi^{2}_{B}$ statistic (L17, Appendix C.1)
for correlated measurement errors.

The need for readily-applied Bayesian model checking procedures 
has recently been persuasively ponted out by Ford - see Sect.4.1 in 
Fischer et al. (2016). Ford attributes the frequent neglect of model 
checking by Bayesians to the absence of a 'neat and tidy' test criterion
for Bayesian analyses, in contrast to the $\chi^{2}$ criterion for
frequentist analyses. For Bayesian model checking, Ford
recommends the book by Gelman et al. (2013), which advocates  
posterior predictive checks. 

In Sect.6.4 of L17, a numerical experiment is briefly reported  
in which $p$-values from posterior predictive checking are compared to those
derived from the same data with the $\chi^{2}_{B}$ statistic. 
Given the importance of establishing
the merits of the simpler approach provided by the $\chi^{2}_{B}$ statistic,
this short paper presents the numerical experiment in some detail.

\section{Bayesian $p$-values}

In this section, the basics of the $p$-values to be compared are stated.

\subsection{Posterior predictive $p$-value}

On the assumption that $\chi^{2}$ is an appropriate test quantity - other
possibilities are discussed in Gelman et al. (2013), the steps required to 
obtain the posterior predictive $p$-value are as follows: \\ 
1) Compute the posterior probability density function, given by
\begin{equation}
 \Lambda(\alpha|D) = \pi(\alpha) {\cal L}(\alpha|D) 
            \: / \int \pi(\alpha) {\cal L}(\alpha|D) dV_{\alpha}   
\end{equation}
where $\pi$ is the prior,
and ${\cal L}$, the likelihood, is given by 
\begin{equation}
 {\cal L} (\alpha|D) = Pr(D|\alpha) \propto  \exp[ - \frac{1}{2} \: 
     \chi^{2}(\alpha|D) ] 
\end{equation}
\\
2) Randomly select a point $\alpha'$ from the posterior density 
$\Lambda(\alpha|D)$.
\\
3) Compute a simulated data set $D'$ at $\alpha'$ by randomly sampling
$Pr(D|\alpha')$.\\
4) Compute and record the values of $\chi^{2}(\alpha'|D')$ and
$\chi^{2}(\alpha'|D)$.\\
5) Repeat steps 2) to 5) ${\cal N}_{tot}$ times.

The resulting Bayesian $p$-value is then
\begin{equation}
       p_{B} = {\cal N} ( \chi^{2} (\alpha'|D') > \chi^{2} (\alpha'|D) )
                         \:/\: {\cal N}_{tot} 
\end{equation}
\subsection{$p$-value from the $\chi^{2}_{B}$ statistic} 
Following earlier work (Lucy 2016; L16) on the statistical properties
of the posterior mean of $\chi^{2}$, the general form of the $\chi^{2}_{B}$ 
statistic is (Appendix C in L17)
\begin{equation}
  \chi^{2}_{B}(D) = \langle \chi^{2} \rangle_{\pi} -k
\end{equation}
where $k$ is the number of parameters and
\begin{equation}
  \langle \chi^{2} \rangle_{\pi} = \int \chi^{2}(\alpha) \Lambda(\alpha|D)
                                               dV_{\alpha}
\end{equation}
The $p$-value corresponding to the $\chi^{2}_{B}$ statistic is derived from
the $\chi^{2}$ distribution with $\nu = n - k$ degrees of freedom, where
$n$ is the number of measurements. Specifically,
\begin{equation}
  p(\chi^{2}_{B}) = Pr(\chi^{2}_{n-k} > \chi^{2}_{B}) 
\end{equation}

The theoretical basis of this $p$-value is as follows:
In the majority of
astronomical applications of Bayesian methods, the priors are weak and 
non-informative. Typically, $\pi(\alpha)$ defines a box in $\alpha$-space
representing a search volume that the investigator is confident 
includes all points $\alpha$ with significant likelihood. In consequence,
a weak prior has neglible effect on the posterior density $\Lambda$ 
and therefore also on any resulting inferences.

In the weak prior limit, Eq.(1) simplifies to 
\begin{equation}
 \Lambda(\alpha|D) =  {\cal L}(\alpha|D) 
            \: / \int  {\cal L}(\alpha|D) dV_{\alpha}   
\end{equation}
and the resulting posterior mean of $\chi^{2}$ is written as
$\langle \chi^{2} \rangle_{u}$.  

Now, in Appendix A of L16, it is proved that, for linearity in 
$\alpha$ and normally-distributed measurement errors, 
\begin{equation}
 \langle \chi^{2} \rangle_{u} - k = \chi^{2}_{0}
\end{equation}
where $\chi^{2}_{0}$ is the minimum value of $\chi^{2}(\alpha|D)$. Since,
under the stated assumptions, the minimum-$\chi^{2}$ solution $\chi^{2}_{0}$
has a $\chi^{2}_{\nu}$  distribution with $\nu = n-k$ degrees of freedom, 
it follows
that, with the extra assumption of a weak prior, the statistic $\chi^{2}_{B}$
defined in Eq.(4) is distributed as $\chi^{2}_{n-k}$. This is the theoretical
basis of the $p$-value given in Eq.(6).

\subsection{$p$-value from the $\psi^{2}_{B}$ statistic} 

The $\chi^{2}_{B}$ statistic of Sect.2.2 is appropriate when the measurement
errors are uncorrelated. When errors are correlated, $\chi^{2}(\alpha|D)$
is replaced by (see Appendix C.1 in L17)
\begin{equation}
 \psi^{2}(\alpha|D) = \vec{v}' \vec{C}^{-1} \vec{v}
\end{equation}
where $\vec{C}$ is the covariance matrix and $\vec{v}$ is the vector of
residuals. The statistic $\psi^{2}$ reduces to $\chi^{2}$ when the 
off-diagonal elements of $\vec{C}^{-1}$ are zero - i.e., no correlations.

By analogy with the statistic $\chi^{2}_{B}(D)$, a statistic 
$\psi^{2}_{B}(D)$ is derived from the posterior mean of $\psi^{2}$.
Specifically, 
\begin{equation}
 \psi^{2}_{B}(D) = \langle \psi^{2} \rangle_{\pi} -k
\end{equation}
with corresponding $p$-value
\begin{equation}
 p_(\psi^{2}_{B}) = Pr(\chi^{2}_{n-k} > \psi^{2}_{B})
\end{equation}

The theoretical basis for this $p$-value is essentially identical to that
for the $p$-value derived from $\chi^{2}_{B}$. If we have linearity in 
$\alpha$ and
normally-distributed errors, then in the weak prior limit
\begin{equation}
  \langle \psi^{2} \rangle_{u} - k = \psi^{2}_{0}
\end{equation}
and $\psi^{2}_{0}$ is distributed as $\chi^{2}_{\nu}$ with $\nu = n-k$
degrees of freedom. 

Note that posterior predictive $p$-values (Sect.2.1) can similarly be 
generalized
to treat correlated measurement errors. The quantity $\psi^{2}$ replaces 
$\chi^{2}$, and the random sampling to create simulated data sets can be
accomplished with a Cholesky decompostion of the covariance matrix
(e.g., Gentle 2009).  
\section{Test model} 
A simple 1-D test model is now defined. This allows numerous
simulated data sets to be created so that 
the $p$-values computed according to Sects.2.1 and 2.2 can be compared
and any differeces established with high statistical confidence.
\subsection{Hubble expansion}
The null hypothesis is that the local Universe is undergoing an isotropic,
cold Hubble expansion with Hubble parameter $h_{0} = 70$ km/s/Mpc.
Moreover throughout the local Universe there exists a population of 
{\em perfect}
standard candles of absolute bolometric magnitude $M = -19.0$. Finally, we 
suppose 
that the redshifts $z$ of these standard candles are measured exactly, but 
that their apparent bolometric
magnitudes $m$ have normally-distributed measurement errors with
$\sigma_{m} = 0.3$ mag.
\subsection{Ideal samples}
In order to test $p$-values when the null hypothesis is correct, we need
data sets consistent with the above description of the Hubble flow. 

We define the local Universe as extending to $z_{max}=0.2$, so that to a good 
approximation the geometry is Euclidean. If we suppose the standard candles
to be uniformly distributed in space, a sample $\{z_{i}\}$ 
is created with the formula
\begin{equation} 
  z_{i} = z_{max} \: x_{i}^{1/3} \;\; for \;\; i=1,2,\dots,n
\end{equation}
where the $x_{i}$ are independent random numbers $\in (0,1)$.
The distances of these $n$ standard candles are
\begin{equation} 
  d_{i} = c z_{i}/h_{0}      \;\;\;  Mpc
\end{equation}
and their measured apparent bolometric magnitudes are
\begin{equation} 
  \widetilde{m}_{i} = M + 5 \log d_{i} + 25 + \sigma_{m} z_{G}
\end{equation}
where the $z_{G}$ are independent gausssian variates from ${\cal N}(0,1)$.

The $n$ pairs $(\widetilde{m}_{i},z_{i})$ comprise a data set $D$ from which 
the posterior probability
density $\Lambda(h|D)$ of the Hubble parameter $h$ can be inferred.
Moreover, the global GOF between  $\Lambda(h|D)$ and $D$ can be tested
with the $p$-values of Sects. 2.1 and 2.2.  
\subsection{Imperfect samples}
In order to test $p$-values when the null hypothesis is false, we need
to modify either the model or the data. If we continue to
make inferences based on the Hubble flow defined in Sect.3.1, we
can introduce a violation of the null hypothesis by contaminating the data
sample with an unrecgnized second class of standard candles
with a different absolute bolometric magnitude.
Thus the data set D now comprises $n_{1}$ pairs $(\widetilde{m}_{i},z_{i})$ 
with $\widetilde{m}_{i}$
given by Eq.(15) and $n_{2}$ pairs $(\widetilde{m}_{i},z_{i})$ with 
$\widetilde{m}_{i}$ given by
\begin{equation} 
  \widetilde{m}_{i} = M - \Delta M + 5 \log d_{i} + 25 + \sigma_{m} z_{G}
\end{equation}
and where $n = n_{1} + n_{2}$. Thus, the sample
is now contaminated with a second population of standard candles that are 
brighter by $\Delta M$ magnitudes. The investigator is unaware of this,
and so carries out a Bayesian analysis assuming an ideal sample.    
\section{Comparison of $p$-values}  
The Bayesian $p$-values defined in Sect.2 are now computed
for simulated data samples derived as described in Sect.3. 
\subsection{Null hypothesis correct}
In order to compare $p$-values in this case, $J$ independent ideal samples 
$D_{j}$ are created as described in Sect.3.2, and then $p$-values for each 
$D_{j}$ computed as described in Sects. 2.1 and 2.2. 
 
With a constant prior $\pi(h)$ in the interval $(h_{1},h_{2})$
with $h_{1} = 64$ and $h_{2} = 76$ km/s/Mpc, the posterior density 
$\Lambda(h|D_{j})$ is obtained from Eq.(1) with the scalar $h$ replacing
the vector $\alpha$. 

The posterior predictive $p$-value for $D_{j}$ is computed as described in
Sect.2.1. First, a value $h'$ is derived by randomly sampling 
$\Lambda(h,D_{j})$. This is achieved by solving the equation
\begin{equation} 
  \int_{h_{1}}^{h'} \Lambda(h,D_{j}) \: dh = x_{\ell} 
\end{equation}
where $x_{\ell}$ is a random number $\in (0,1)$. Second, an ideal 
sample $D'$ for
Hubble parameter $h'$ is computed according to Sect.3.2. Third, the 
GOF values $\chi^{2}(h'|D')$ and $\chi^{2}(h'|D_{j})$
are computed, where
\begin{equation} 
  \chi^{2}(h,D) = \sum_{1}^{n} (\widetilde{m}_{i} - m_{i})^{2}/ \sigma_{m}^{2}  
\end{equation}
where the predicted apparent bolometric magnitude is
\begin{equation} 
  m_{i} = M + 5 \log cz_{i}/h + 25 
\end{equation}

These steps are repeated ${\cal N}_{tot}$ times with independent random numbers
$x_{\ell}$ in Eq.(17). The posterior predictive $p$-value for the ideal sample
$D_{j}$ is then given by Eq.(3).

The corresponding $p$-value for $D_{j}$ from the $\chi^{2}_{B}$ statistic
is given by Eq.(4), where
\begin{equation} 
  \langle \chi^{2} \rangle_{\pi} = \int \chi^{2}(h,D_{j}) \: 
                                 \Lambda(h|D_{j}) \: dh 
\end{equation}
with $\chi^{2} (h,D)$ from Eq.(18).
\subsubsection{Numerical results}

The calculations described in Sects.4.1 and 4.2 is carried out for $J = 200$ 
ideal samples $D_{j}$, with $D_{j}$ comprising $n = 200$ pairs 
$(z_{i},\widetilde{m}_{i})$.
In calculating posterior predictive $p$-values, we take 
${\cal N}_{tot} = 10^{5}$ randomly selected Hubble parameters $h'$
for each $D_{j}$ in order to achieve high precision for $p_{B}$.  

The results are plotted in Fig.1. This reveals superb agreement between the
two $p$-values with no outliers. The mean value of 
$|\Delta log p| = 2.3 \times 10^{-3}$ confirms that there is almost exact
agreement. Accordingly, this experiment indicates that the 
readily-calculated $p(\chi^{2}_{B})$
is an almost exact proxy for the posterior predictive 
$p$-value $p_{B}$,
the calculation of which is undoubtedly cumbersome.
\begin{figure}
\vspace{8.2cm}
\includegraphics{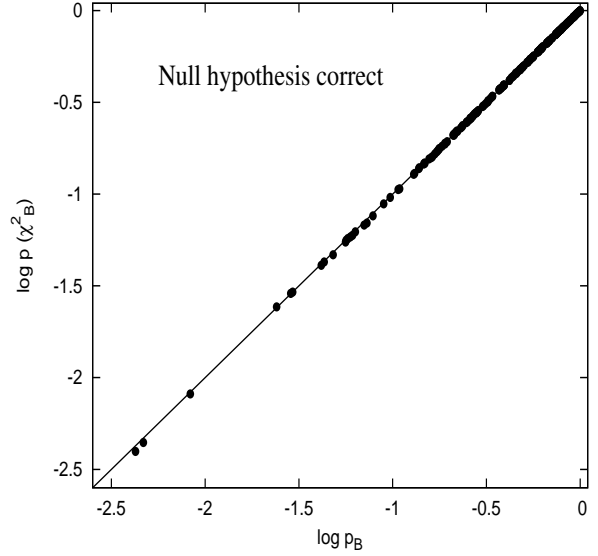}
\caption{Comparison of $p$-values. For $J = 200$ ideal samples $D_{j}$, 
$\log p (\chi^{2}_{B})$ is plotted against $\log p_{B}$.} 
\end{figure}

\subsection{Null hypothesis false}

In order to compare $p$-values when $p \la 0.001$, $J = 200$ independent 
{\em imperfect} samples $D_{j}$ are created as described in Sect.3.3. Again 
${\cal N}_{tot} = 10^{5}$ and $n = 200$, but now $n = n_{1} + n_{2}$, with
$n_{1} = 180$ perfect standard candles and $n_{2} = 20$ belonging to a second 
population brighter by $\Delta M = 0.5$mag. With these changes, the calculation
of Sect.4.1.1 is repeated.

\subsubsection{Numerical results}    

The results are plotted in Fig.2. Because the null hypothesis is false,
$p$-values $< 0.01$ occur frequently, so that the comparison provided by
Fig.1 now extends to $p \la 10^{-4}$. However, at these small $p$-values,
the values of $p_{B}$ are subject to noticable poisson noise even with 
${\cal N}_{tot} = 10^{5}$. Because of this,  
$|\Delta log p| = 3.2 \times 10^{-2}$, somewhat larger than in Sect.4.1.1
but neverthless again confirming the excellent agreement between the
two $p$-values.

\begin{figure}
\vspace{8.2cm}
\includegraphics{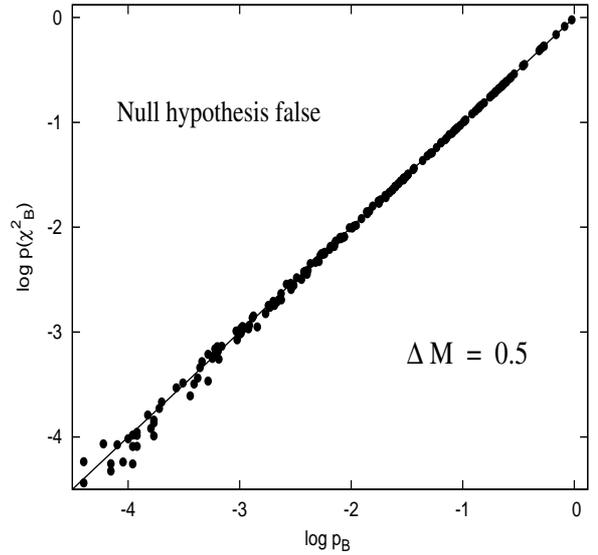}
\caption{Comparison of $p$-values for imperfect samples. Each imperfect sample
contains 180 standard candles of magnitude $M$ and 20 of magnitude 
$M - \Delta M$ with $\Delta M =0.5$.}
\end{figure}

In Fig.3, the behavior of the $p$-values derived from the $\chi^{2}_{B}$
statistic as a function of $\Delta M$ is plotted. With increesing 
$\Delta M$, the points shift to smaller $p$-values, increasing the
likelihood that the investigator will conclude that the Bayesian model
is suspect.

\begin{figure}
\vspace{8.2cm}
\includegraphics{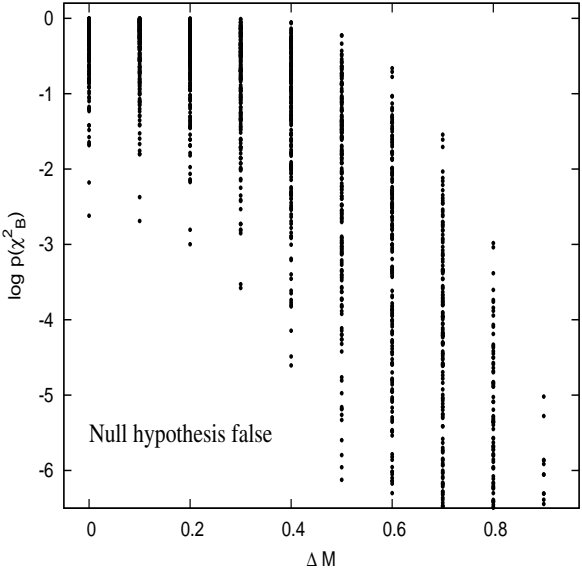}
\caption{$p$-values for imperfect samples. Each imperfect sample contains
180 standard candles of magnitude $M$ and 20 of magnitude $M - \Delta M$.
For each $\Delta M$, the $p$ values are computed for 200 independent samples. } 
\end{figure}
\subsection{Conclusion}

The aim of this paper is to investigate the performance of
a recenly proposed global GOF criterion $\chi^{2}_{B}$ for 
posterior probability densities functions.
To do this, a toy model of the local Hubble expansion is defined (Sect.3)
for which both ideal (Sect.3.2) and flawed (Sect.3.3) data samples can
be created. The $p$-values derived for such samples from $\chi^{2}_{B}$
are then compared to the $p$-values obtained with the well-established 
procedure
of posterior predictive checking. The two $p$-values are found to be in close 
agreement (Figs.1 and 2) throughout the interval (0.0001, 1.0), thus
covering the range of $p$-values expected when a Bayesian model
is supported ($p \ga 0.01$) to those when the model is suspect ($p \la 0.01$).

Because carrying out a posterior predictive check is cumbersome or
even infeasible, the $p$-value derived from $\chi^{2}_{B}$ can serve
as an effective proxy for posterior predictive checking. Of course, this 
recommendation derives from just 
one statistical experiment. Other investigators may wish to devise
experiments relevant to different areas of astronomy.

\end{document}